\documentclass[apj]{emulateapj}

\shorttitle{Two-dimensional hydrodynamical case}

\shortauthors{Bu et al.}

\begin{document}

\title{Hydrodynamical numerical simulation of wind production from black hole hot accretion flows at very large radii}

\author{De-Fu Bu\altaffilmark{1}, Feng Yuan\altaffilmark{1}, Zhao-Ming Gan\altaffilmark{1}, Xiao-Hong Yang\altaffilmark{2}}


\altaffiltext{1}{Key Laboratory for Research in Galaxies and
Cosmology, Shanghai Astronomical Observatory, Chinese Academy of
Sciences, 80 Nandan Road, Shanghai, 200030, China; fyuan@shao.ac.cn}
\altaffiltext{2}{Department of Physics, Chongqing University,
Chongqing 400044, China}

\begin{abstract}
Previous works show strong winds exist in hot accretion flows around
black holes. Those works focus only on the region close to the black
hole thus it is unknown whether or where the wind production stops
at large radii. In this paper, we investigate this problem by
hydrodynamical simulations. We take into account gravities of both
the black hole and the nuclear star clusters. For the latter, we
assume that the velocity dispersion of stars is a constant and its
gravitational potential $\propto \sigma^2 \ln (r)$, where $\sigma$
is the velocity dispersion of stars and $r$ is the distance from the
center of the galaxy. We focus on the region where the gravitational
potential is dominated by the star cluster. We find, same as the
accretion flow at small radii, the mass inflow rate decreases inward
and the flow is convectively unstable. However, trajectory analysis
shows that there is very few wind launched from the flow. Our
result, combined with the results of Yuan et al. (2015), indicates
that the mass flux of wind launched from hot accretion flow
$\dot{M}_{\rm wind}=\dot{M}_{\rm BH}(r/20r_s)$, with $r\la R_A\equiv
GM_{\rm BH}/\sigma^2$. Here $\dot{M}_{\rm BH}$ is accretion rate at
black hole horizon. $R_A$ is similar to Bondi radius. We argue that
the inward decrease of inflow rate is not because of mass loss via
wind, but because of convective motion. The disappearance of wind
outside of $R_A$ must be because of the change of the gravitational
potential, but the exact reason remains to be probed.
\end{abstract}

\keywords {accretion, accretion disks $-$ black hole physics $-$
hydrodynamics}

\section{INTRODUCTION}

Hot accretion flow such as advection-dominated accretion flow
(ADAFs; Narayan \& Yi 1994; 1995; Abramowicz et al. 1995) is an
important class of accretion mode because it is now believed to be
the standard model of low-luminosity active galactic nuclei, which
are the majority of galaxies at least in the nearby universe, and
the hard and quiescent states of black hole X-ray binaries (see Yuan
\& Narayan 2014 for the most updated review of our current theoretical understanding of hot
accretion flow and its various astrophysical applications).

One of the most important progresses in this filed in recent years
is the finding of the strong wind (real outflow) launched
from the accretion flow. The study starts
from the pioneer works by Stone et al. (1999, hereafter SPB99) and
Igumenshchev \& Abramowicz (1999). By performing two-dimensional
global hydrodynamical (HD) simulations, they have found that the
mass inflow rate (see eq. \ref{inflowrate} for definition) decreases
with decreasing radius. This study has been extended to simulations
with larger radial dynamical range, more spacial dimensions, and
inclusion of magnetic field  (e.g., Stone \& Pringle 2001; Hawley \&
Balbus 2002; Machida et al. 2001; Pang et al. 2011; Yuan et al.
2012a). In general, it is found that $\dot{M}_{\rm in}(r)\propto
r^s$ with $s \sim 0.5-1$ (see a short review in Yuan et al. 2012a).
Yuan et al. (2012b, hereafter YBW12) show that the inward decrease of the
accretion rate is due to the significant mass loss via wind (see
also Narayan et al. 2012; Li et al. 2013; Gu 2015). This conclusion
is soon confirmed by the 3 million seconds {\it Chandra} observation
to the accretion flow around the super-massive black hole in the
Galactic center, combined with the modeling to the detected iron
emission lines (Wang et al. 2013).

By performing the trajectory analysis of virtual test particles
in the accretion flow based on the three dimensional general
relativistic magneto-hydrodynamic (GRMHD) simulation data of
accretion flows, recently Yuan et al. (2015) have analyzed the motion of
winds and calculated their various physical properties. The main findings
are as follows. Firstly, for a non-rotating black hole, the mass
flux of the wind can be described by $\dot{M}_{\rm wind}\approx
\dot{M}_{\rm BH}(r/20r_s)$, with $\dot{M}_{\rm BH}$ is the mass
accretion rate at the black hole horizon and $r_s$ is the
Schwarzschild radius. Secondly, the poloidal speed of wind
originating from radius $r$ is $v_p(r)\approx (0.2-0.4) v_K (r)$
where $v_K(r)$ is the Keplerian speed at radius $r$. The poloidal
speed of wind   remains roughly unchanged during their outward
motion. This is because the centrifugal force and gradient of gas
and magnetic pressure can do work to the wind and this work can
compensate the gravitational work. Also because of this reason, the
initial Bernoulli parameter of wind needs not to be positive. At
last, Yuan et al. (2015) find that close to the rotation axis, with
$\theta \la 15^{\circ}$ in the spherical coordinate, the poloidal
speed of wind is quasi-relativistic, much higher than that of the
other part of wind. This kind of outflow is identified as the
``disk-jet'' (Yuan \& Narayan 2014), which is to discriminate it
from the Poynting-flux dominated ``Blandford-Znajek'' jet. The
``disk-jet'' or the ``BZ-jet'', which one corresponds to the
observed radio jet remains unclear. All these properties of wind are
essential to the study of AGN feedback since wind plays an important
role in the feedback process (e.g., Ostriker et al. 2010).

\begin{figure}
\epsscale{1.2} \plotone{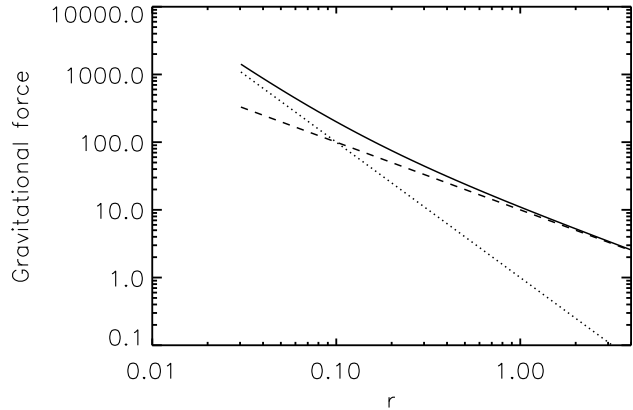}\hspace{1.cm} \epsscale{1.8}
\caption{The distribution of gravitational force. The dashed and
dotted lines correspond to the gravitational force of nuclear star
cluster and black hole, respectively. The solid line is their sum. }
\label{Fig:gravity}
\end{figure}

From the equation  $\dot{M}_{\rm wind}=\dot{M}_{\rm BH}(r/20r_s)$,
we know that most of the wind comes from the region of large radius.
Then a question is how large the value of $r$ can be in the above
equation.  All numerical
simulations of accretion wind so far focus only on the region
relatively close to the black hole, so it is difficult for them to
answer this question. To answer this question, we have to study the
accretion flow far away from the black hole. In this case, the main
change is the gravitational potential of the system. In this case,
in addition to the potential of the black hole, the gravitational
potential  of the nuclear star cluster will become important and
should be included. We define a radius $R_{A}$ at which  the gravitational force
due to the central black hole is equal to that due to the nuclear
star cluster. We call $R_A$ to be the boundary of the accretion flow
or active galactic nuclei (AGNs). Around $R_A$, both the depth and
slope of the gravitational potential will change compared to the
case of a pure black hole potential (see Figure \ref{Fig:gravity}).
Such a change will potentially change the dynamics of both accretion
and wind production. Hereafter, for convenience, we use BHAF to
refer the hot accretion flow close to the black hole, and CAAF
(circum-AGN accretion flow) to refer the hot accretion flow at large radii, around
the boundary of an AGN.

We use numerical simulation to study the wind production in a CAAF.
In this paper we do not include magnetic field and focus only on
hydrodynamical case. We introduce an anomalous stress to transfer
the angular momentum. In reality,  it is the magneto-hydrodynamic
(MHD) turbulence associated with the magneto-rotational instability
(MRI; Balbus \& Hawley 1991, 1998) that is responsible for the
angular momentum transfer in accretion flows. So MHD simulation is
more realistic, which will be our future work. One question of
performing HD simulation is then whether the results are realistic.
This question has been addressed in YBW12 and Yuan et al. (2015). In
YBW12, they have shown that although winds exist in both HD and MHD
accretion flows, the mechanisms of wind production are different. In
the HD case, the accretion flow is convectively unstable; winds are
accelerated by the buoyant force associated with convection, i.e,
the gradient of gas pressure. But in the MHD case, the flow is
convectively stable; winds are accelerated by the combination of
centrifugal force and the gradient of gas and magnetic pressure
(Yuan et al. 2015). Even though these differences, the presence of
wind seems to be irrelevant to the presence of magnetic field
(Begelman 2012). The reason is as follows. In hot accretion flow,
radiation can be neglected. Then there are three energy fluxes in
the accretion flow. The first flux is the liberation of effective
gravitational potential energy. The second is the thermal energy
flux transferred by the motion of gas in the accretion flow. The
third one is the energy flux transferred by viscosity. If we
integrate the energy equation we will have $\dot M Be+T\Omega=Q$. In
this equation, $\dot M$ is the mass accretion rate, $Be$ is the
Bernoulli parameter, $T$ is the viscous stress per unit radius,
$\Omega$ is the angular velocity. $\dot M Be$ corresponds to the
first and second energy fluxes, $T\Omega$ is the energy flux
transferred by viscosity, $Q$ is the net energy flux which is not
zero. For a steady state, the net energy flux should be a constant
of radius. Therefore, $\dot M Be$ and $T\Omega$ should also be
constant of radius. Because $Be$ is proportional to $1/r$,
 $\dot M $ is required to be proportional to $r$.
Therefore, basically, it is required that $\dot M $ is proportional
to $r$ independent of whether the model is hydrodynamic or MHD.
Thus, we believe that our result on the wind production based on HD
simulation should be same with the MHD case. In addition,
academically HD simulation should be useful for us to understand the
MHD simulation result.

The structure of the paper is arranged as follows. In \S2, we will describe
the basic equations and the simulation method. In \S3, we will first
study the convective stability of CAAFs, then calculate the radial
profile of the inflow rate. We will then study whether wind can be
produced in CAAFs using the trajectory method. We discuss and
summarize our results in \S4.

\section{method}\label{method}
\subsection{Equations}

In a spherical coordinate $(r, \theta, \phi)$, we solve the following
hydrodynamical equations describing accretion:
\begin{equation}
\frac{d\rho}{dt}+\rho\nabla\cdot \mathbf{v}=0,\label{cont}
\end{equation}
\begin{equation}
\rho\frac{d\mathbf{v}}{dt}=-\nabla p-\rho\nabla
\psi+\nabla\cdot\mathbf{T}, \label{monentum}
\end{equation}
\begin{equation}
\rho\frac{d(e/\rho)}{dt}=-p\nabla\cdot\mathbf{v}+\mathbf{T}^2/\mu.
\label{energyequation}
\end{equation}
Here, $\rho$, $p$, $\mathbf{v}$, $\psi$, $e$ and $\mathbf{T}$ are
density, pressure, velocity, gravitational potential, internal
energy and anomalous stress tensor, respectively; $d/dt(\equiv
\partial / \partial t+ \mathbf{v} \cdot \nabla)$ denotes the Lagrangian
time derivative. We adopt an equation of state of ideal gas
$p=(\gamma -1)e$, and set $\gamma =5/3$.

The gravitational potential $\psi$ can be expressed as
\begin{equation}
\psi= \psi_{BH}+\psi_{star}.
\end{equation}
The black hole potential $\psi_{BH}=-GM_{BH}/(r-r_s)$, where $G$ is
the gravitational constant, $M_{BH}$ is the mass of the black hole
and $r_s$ is the Schwarzschild radius.  We assume that the velocity dispersion
of nuclear stars is a constant of radius. This seems to be the case of many AGNs. So the potential
of the star cluster is $\psi_{star}=\sigma^2 \ln (r)+C$, where
$\sigma$ is the velocity dispersion of stars and $C$ is a constant. So we have
\begin{equation}
R_A=GM_{\rm BH}/\sigma^2.
\end{equation} We set $\sigma^2=10$ and $G=M_{\rm BH}=1$ to define our units in the present work. So we have $R_A=0.1$.    Fig. 1 shows the gravitational force distribution.

We use the stress tensor $\mathbf {T}$ to mimic the shear stress.
Following SPB99, we assume that the only non-zero components of
$\mathbf {T}$ are the azimuthal components,
\begin{equation}
  T_{r\phi} = \mu r \frac{\partial}{\partial r}
    \left( \frac{v_{\phi}}{r} \right),
    \label{shearviscosity}
\end{equation}
\begin{equation}
  T_{\theta\phi} = \frac{\mu \sin \theta}{r} \frac{\partial}{\partial
  \theta} \left( \frac{v_{\phi}}{\sin \theta} \right) .
\end{equation}
This is because the MRI is driven only by the shear associated with
orbital dynamics. Other components of the stress are much smaller
than the azimuthal components (Stone \& Pringle 2001). We adopt the
coefficient of shear viscosity $\mu=\nu\rho$. In models A1, A2, A4
and B, we assume $\nu\propto r^{1/2}$, which is the usual
``$\alpha$'' viscosity description (SPB99). In order to study the
dependence of results on shear stress, in model A3, we assume
$\nu\propto r$. In all models, we set $\alpha$=0.01 to eliminate the
discrepancy due to the magnitude of viscosity coefficient.

\subsection{Initial conditions}
As for the initial density distribution, we assume a rotating
equilibrium torus embedded in a non-rotating, low-density medium. We
assume that the gas torus has constant specific angular momentum $L$
and assume a polytropic equation of state, $p=A\rho^\gamma$, where
$A$ is a constant. Using these assumptions, we can integrate the
equation of motion into the potential form (Nishikori et al. 2006)
\begin{equation}
\Psi(r, \theta)=\psi+\frac{L^2}{2(r
\sin\theta)^2}+\frac{\gamma}{\gamma-1}\frac{p}{\rho}
 =\Psi(R_0,\pi/2),
\end{equation}
where $R_0$ is the radius of density maximum  of the torus. Using
equation (7), we obtain the density distribution,
\begin{equation}
\rho=\rho_c \left \{\frac{\max[\Psi(R_0,\pi/2)-\psi(r,
\theta)-L^2/(2(r
\sin\theta)^2),0]}{A[\gamma/(\gamma-1)]}\right\}^{1/(\gamma-1)}
\end{equation}
where $\rho_c$ is the density at the torus center. In this paper, we
assume $\rho_c=1$ and $A=0.4$.

The ambient medium in which the torus is embedded has density
$\rho_0$ and pressure $\rho_0/r$. We choose $\rho_0=10^{-4}$. The
mass and pressure of the ambient medium is negligibly small.

\subsection{Models}
In this paper, we simulate both CAAFs and BHAFs. Models A1-A4 are
for CAAFs. In models A1 and A2, we assume viscosity $\nu\propto
r^{1/2}$. The only difference between models A1 and A2 is that the
resolution of model A2 is two times of that of model A1. In model
A3, we set $\nu\propto r$. The purpose to carry out model A3 is to
study the dependence of results on viscosity form. To study the
boundary effect, we carry out model A4. In model A4, the
computational domain is $0.002<r<0.4$. In models A1, A2, and A3, the
center of the initial torus is located at $r=1$. In model A4, the
center is located at $r=0.1$ In model B, we simulate a BHAF. Table 1
summarizes all these models.

\begin{table*}
\footnotesize
\begin{center}
\caption{models in this paper}
\begin{tabular}{ccccc} \\ \hline
Models & Shear stress & Resolution  &  Computational domain  & Gravitational potential\\
\hline

A1 &  $ \nu \propto r^{1/2}$    & $334 \times 160$  & $0.02 \leq r \leq 4$ & Star cluster gravity + Black hole gravity\\
A2 & $ \nu \propto r^{1/2}$    & $668 \times 320$ & $0.02 \leq r \leq 4$ & Star cluster gravity + Black hole gravity\\
A3 &  $ \nu \propto r      $    & $334 \times 160$  & $0.02 \leq r \leq 4$ & Star cluster gravity + Black hole gravity\\
A4 &  $ \nu \propto r^{1/2}$    & $334 \times 160$  & $0.002 \leq r \leq 0.4$ & Star cluster gravity + Black hole gravity\\
B  &  $ \nu \propto r^{1/2}$    & $334 \times 160$  & $2r_s \leq r \leq 400r_s$ &   Black hole gravity\\
\hline
\end{tabular}\\
\end{center}
\end{table*}

\subsection{Numerical Method}
We use the ZEUS-2D code (Stone \& Norman 1992a,1992b) to solve
equations (1)-(3). The polar angle range is $0 \leq \theta \leq
\pi$. We adopt non-uniform grid in the radial direction
$(\bigtriangleup r)_{i+1} / (\bigtriangleup r)_{i} = 1.037$. The
distributions of grids in $\theta$ direction in the northern and
southern hemispheres are symmetric about the equatorial plane. The
resolution at $\theta$ in the northern hemisphere is same as that at
$\pi-\theta$ in the southern hemisphere.In order to well resolve the
accretion disk around the equatorial plane, the resolution is
increased from the north and south rotational axis to the equatorial
plane with $(\bigtriangleup \theta)_{j+1} / (\bigtriangleup
\theta)_{j} = 0.9826$ for $0 \leq \theta \leq \pi/2$ and
$(\bigtriangleup \theta)_{j+1} / (\bigtriangleup \theta)_{j} =
1.0177$ for $\pi/2 \leq \theta \leq \pi$.  At the poles, we use
axisymmetric boundary conditions. At the inner and outer radial
boundary, we use outflow boundary conditions.

\section{RESULTS}\label{results}

\subsection{Convective stability}
\begin{figure}
\epsscale{1.2} \plotone{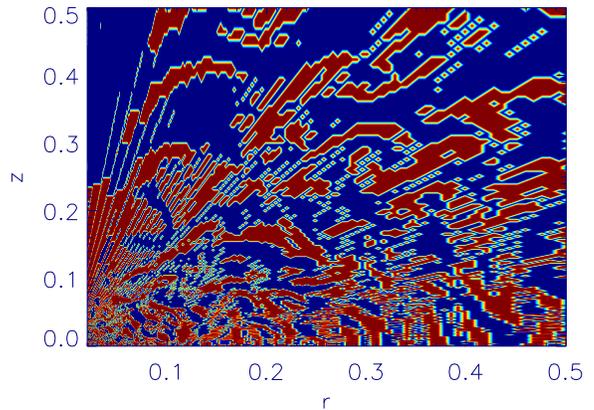}\hspace{1.cm} \epsscale{1.8}
\caption{Convective stability analysis of Model A1. The result is
obtained according to eq. (\ref{hoiland}) based on the simulation
data at t=13.2 orbits at the initial torus center $r=1$. The red
region is unstable.} \label{Fig:instability}
\end{figure}

\begin{figure*}
\includegraphics[width=18cm]{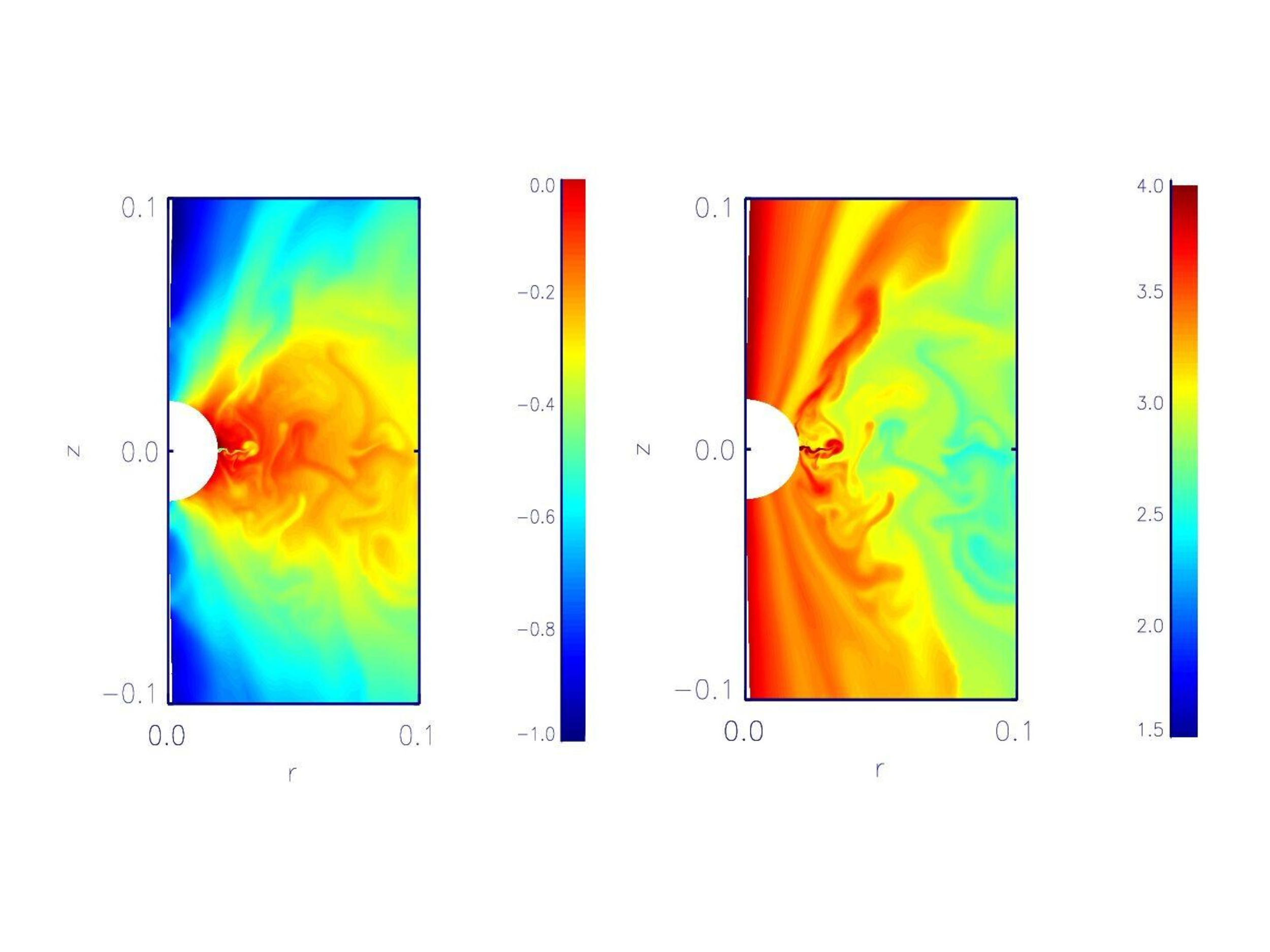}
\vspace{-2cm}
\caption{Images of the logarithm density (left-panel) and entropy
(right-panel) for model A1 at time $t=12.7$ orbits. Note the correlation between regions of high (low)
density and low (high) entropy near the equatorial plane.} \label{Fig:snapshot}
\end{figure*}

HD simulations of BHAFs  (e.g., SPB99; Igumenshchev \& Abramowicz
1999, 2000; Yuan \& Bu 2010) have found that the flows are convectively
unstable, consistent with what has been suggested by the
one-dimensional analytical study of BHAFs (Narayan \& Yi 1994). The
physical reason is that the entropy of the flow increases inward,
which is resulted by the viscous heating and negligible radiative
loss. As we state in \S1, YBW12 shows that strong winds are produced
in HD accretion flow around a black hole. The force driving the wind
is mainly the buoyant force associated with the convective
instability.

To study  whether the CAAF is convectively stable or not, we use the
H{\o}iland criteria (e.g., Tassoul 1978; Begelman \& Meier 1982):
\begin {equation}(\nabla s \cdot \mathbf{dr})(\mathbf{g} \cdot
\mathbf{dr})-\frac{2\gamma v_{\phi}}{R^2}[\nabla(v_{\phi}R)\cdot
\mathbf{dr}]dR <0.\label{hoiland}\end{equation}  In equation
(\ref{hoiland}), $R=r\sin \theta$ is the cylindrical radius, ${\bf
dr} = dr \hat r + r d \theta \hat \theta$ is the displacement
vector, $s = \ln(p) - \gamma \ln(\rho)$ is $(\gamma - 1)$ times the
entropy, ${\bf g} = - \nabla \psi + \hat R v_{\phi}^2/R$ is the
effective gravity, and $v_{\phi}$ is the rotational velocities. For
non-rotating flow, this condition is equivalent to an inward
increase of entropy, which is the well-known Schwarzschild criteria.

Taking model A1 as an example, Figure \ref{Fig:instability} shows
the result. The result is obtained according to eq. (\ref{hoiland})
based on the simulation data at t=13.2 orbits at the initial torus
center $r=1$ (here and hereafter, we use the orbital period at the
center of initial torus as the time unit). At $t=13.2$ orbits, the
flow has achieved a steady state since the net accretion rate
averaged between $t=12.7$ to 15 orbits is a constant of radius (see
figure \ref{Fig:accretionrate}, the dotted line). The red regions
are convectively unstable. So we can see that CAAF is mainly
convectively unstable. The physical reason for the instability is
same as for the BHAF. That is, during the accretion process, viscous
heating produces entropy, while the loss of entropy by radiative
cooling is neglected.  Figure \ref{Fig:snapshot} shows the snapshot
for logarithm density (left-panel) and entropy $S=\ln
(p/\rho^\gamma)$ (right panel) for model A1 at time $t=12.7$ orbits.
From the poles to the equator, the density is strongly stratified.
On small scales, large-amplitude fluctuations associated with
convection dominate. The entropy plot shows that entropy is maximum
along the poles and smallest along the equator. On small scales,
bubbles and filaments associated with convection dominate the image.
Comparison of the two images demonstrates that regions of high
density have low entropy, while bubbles of low density have high
entropy. This is the typical feature of convection.

\subsection{Mass inflow rate}
Following SPB99, we define the mass inflow and outflow rates, $\dot
{M}_{\rm in}$ and $\dot {M}_{\rm out}$, as follows,

\begin{equation}
 \dot{M}_{\rm in}(r) = 2\pi r^{2} \int_{0}^{\pi} \rho \min(v_{r},0)
   \sin \theta d\theta
   \label{inflowrate}
\end{equation}
\begin{equation}
 \dot{M}_{\rm out}(r) = 2\pi r^{2} \int_{0}^{\pi} \rho \max(v_{r},0)
    \sin \theta d\theta
    \label{outflowrate}
\end{equation}
The net mass accretion rate is,
\begin{equation}
\dot{M}_{\rm acc}(r)=\dot{M}_{\rm in}(r)+\dot{M}_{\rm out}(r)
\label{netrate}\end{equation} Note that the above rates are obtained
by time-averaging the integrals rather than integrating the time
averages.

\begin{figure}
\includegraphics[width=8cm]{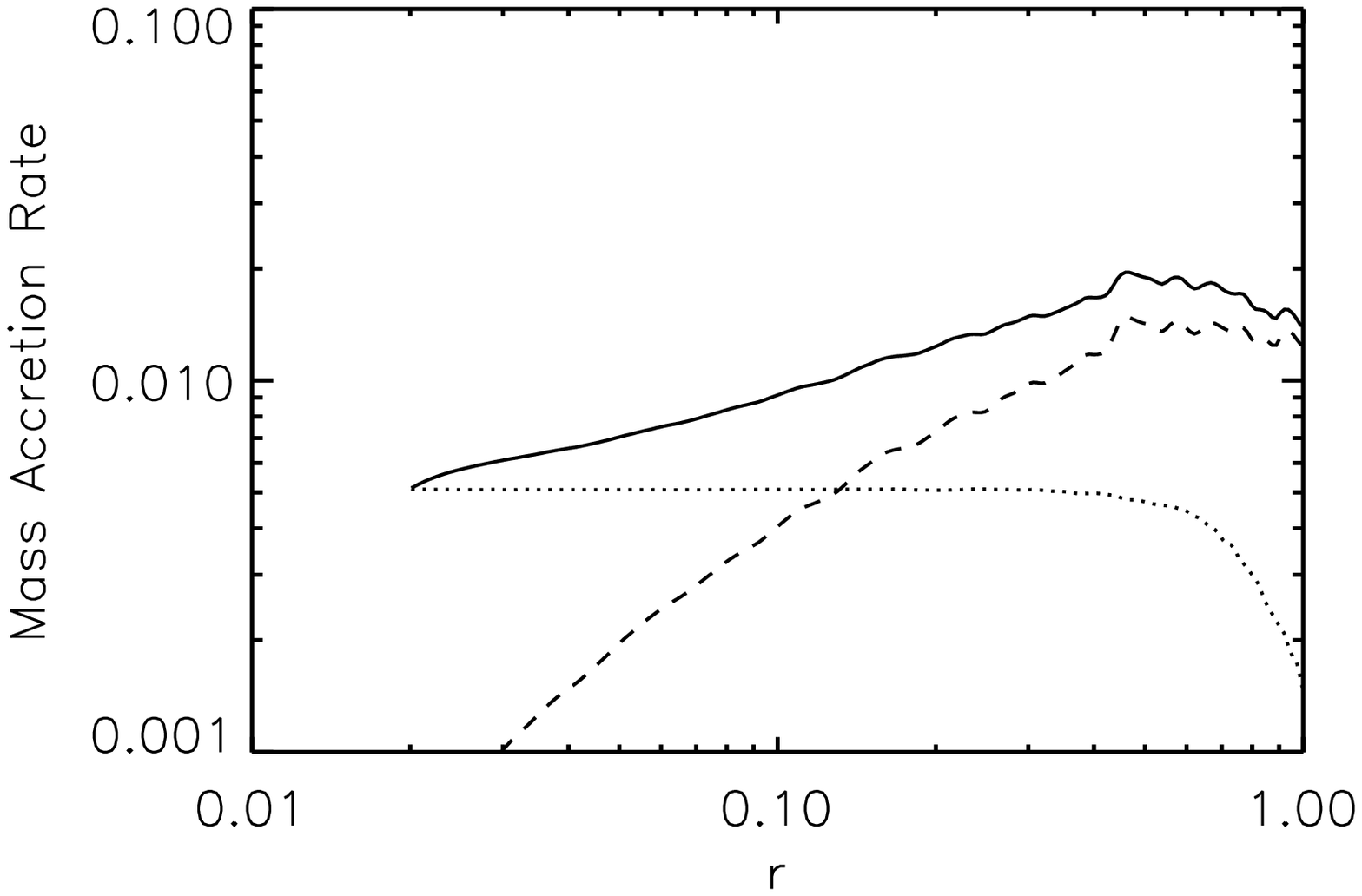}
\caption{The radial profiles of the time-averaged (from $t=12.7$ to
15 orbits) and angle integrated mass inflow rate $\dot{M}_{\rm in}$
(solid line), outflow rate $\dot{M}_{\rm out}$ (dashed line), and
the net rate $\dot{M}_{\rm acc}$ (dotted line) in model A1.  \label{Fig:accretionrate}}
\end{figure}

\begin{figure}
\includegraphics[width=8cm]{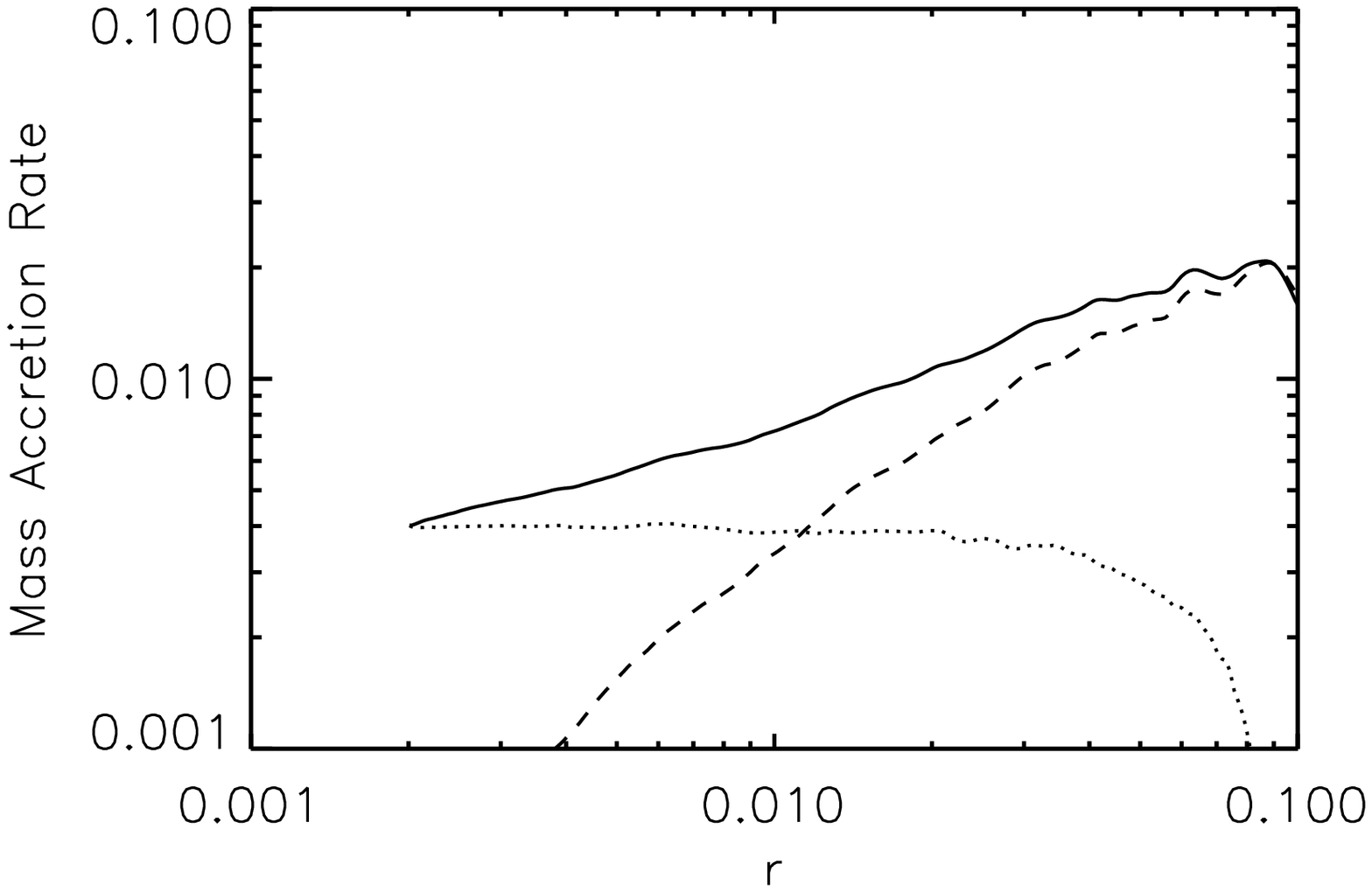}
\caption{The radial profile of the time-averaged (from $t=4.5$ to
6.8 orbits) and angle integrated mass inflow rate $\dot{M}_{\rm in}$
(solid line), outflow rate $\dot{M}_{\rm out}$ (dashed line), and
the net rate $\dot{M}_{\rm acc}$ (dotted line) in model A4. \label{Fig:accretionratemovein}}
\end{figure}

\begin{figure}
\includegraphics[width=14cm]{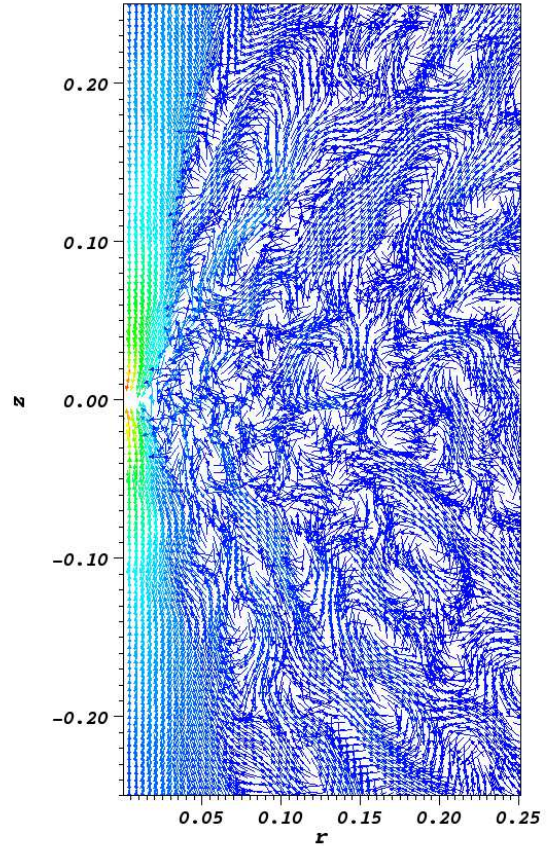}
\vspace{-0.7cm} \caption{Snapshot of velocity vector of model A1 at
$t=12.7$ orbits. Convective eddies occupy the whole domain and it is hard
to find winds.} \label{Fig:vector}
\end{figure}

\begin{figure}
\includegraphics[width=8cm]{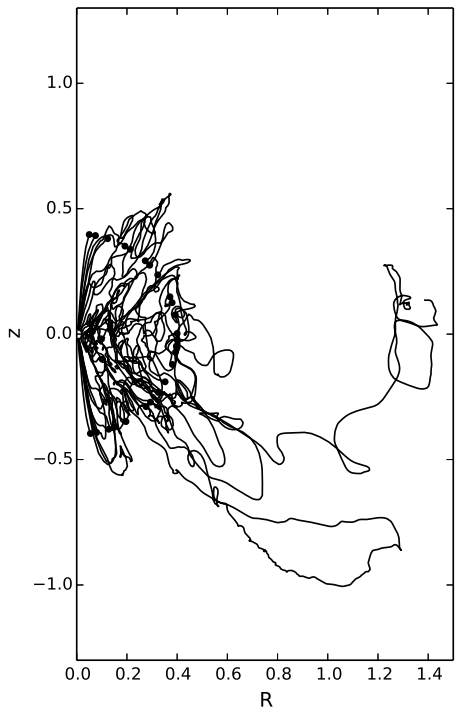}
\caption{Trajectory of gas for model A1. The black dots located at
$r=0.4$ is the starting points of the test particles. Winds are very
weak. \label{Fig:trajectory}}
\end{figure}

Figure \ref{Fig:accretionrate} shows the time-averaged (from 12.7-15
orbits) and angle-integrated mass accretion rate in model A1. The
radial profile of the inflow rate from $r=0.02$ to $0.4$ can be
described by
\begin{equation}
\dot M_{\rm in} \propto r^{0.42}. \end {equation} As a comparison,
HD simulations of BHAF with a large radial dynamical range by Yuan
et al. (2012a) find $\dot M_{\rm in} \propto r^{0.54}$ in the model
with the same value of $\alpha$.

In the region $0.02<r<0.1$, the outflow rate is much smaller than
the mass inflow rate. This region is dominated by the gravity of
black hole. Previous works (Yuan et al. 2012b; Yuan et al. 2015)
have shown that in this case outflow (wind) should be strong. The
apparent discrepancy between this work and our previous works is due
to the fact that the region $r<0.1$ is too close to the inner
boundary where a somewhat ``unphysical'' boundary condition (i.e.,
outflow condition) is adopted. The result will be reliable if we
focus on region far away from the inner boundary. In order to
illustrate this point, we have carried out model A4. In this model,
our computational domain is $0.002< r < 0.4$. The result is shown in
figure \ref{Fig:accretionratemovein}. In the region $0.01 < r < 0.1$
which is far away from the inner boundary, the outflow rate
obviously becomes very high, which is consistent with our previous
works.

\subsection{Does strong wind exist in a CAAF?}

Figure \ref{Fig:accretionrate} shows that there is a significant
mass outflow rate. But this does not mean it is real outflow (wind)
because it may be due to the convective turbulent motion. The
turbulence is induced by convective instability so we call the
turbulence ``convective turbulence''. To study whether wind exists,
let us first directly look at the velocity field shown in Figure
\ref{Fig:vector}. Clearly, convective eddies occupy the whole
domain. From equatorial plane to the pole, the scale of convective
eddies becomes larger. From large to small radii, the scale of
convective eddies becomes smaller. From this figure, it is hard to
find winds.

Following Yuan et al. (2015), we now use the much more precise
trajectory method to study whether winds exist. The details of this
approach can be found in Yuan et al. (2015), here we only briefly
introduce it. Trajectory is obtained by connecting the positions of
the same ``test particle'' at different time. This concept is
related with the Lagrangian description of fluid. Note that this is
different from the streamline which is obtained by connecting the
velocity vector of different test particles with infinitely short
distance at a given time. This concept is associated with the Euler
description of fluid motion. Trajectory is only equivalent to the
streamline for strictly steady motion, which is not the case for
accretion flow since it is always turbulent. To get the trajectory,
we first need to choose some virtual ``test particle'' in the
simulation domain. They are of course not real particles, but some
grids representing fluid elements. Their locations and velocity at a
certain time $t$ are obtained directly from the simulation data. We
can then obtain their location at time $t +\delta ¦Ät$ from the
velocity vector and $\delta t$.
 We do this work using a software
called ``VISIT''.

Using this trajectory approach, we can discriminate easily which
particles are real outflows (i.e., winds) and which are doing
turbulent motions. We can also calculate the mass fluxes of wind and
turbulent outflow by combining with the information of density and
velocity of wind. For details, see Yuan et al. (2015). Figure
\ref{Fig:trajectory} shows the trajectory of 25 ``test particles''
starting from $r=0.4$ in model A1. From this figure we can see that
the real wind trajectories, i.e., the trajectories which extend from
$r=0.4$ to large radius and never come across $r=0.4$ twice,  are
very few. This implies that the mass flux of wind is very small. Our
quantitative calculation confirms this result. For example, we find
that at $r=0.4$, the ratio of mass flux of real outflows to the
total outflow rate calculated by equation (\ref{outflowrate}) is
only $0.3\%$. This result means that there is almost no wind. As a
comparison, in the case of BHAF, Yuan et al. (2015) find  that this
ratio is $\sim 60\%$.

\subsection{Why the inflow rate decreases inward in a CAAF?}
\begin{figure}
\includegraphics[width=8cm]{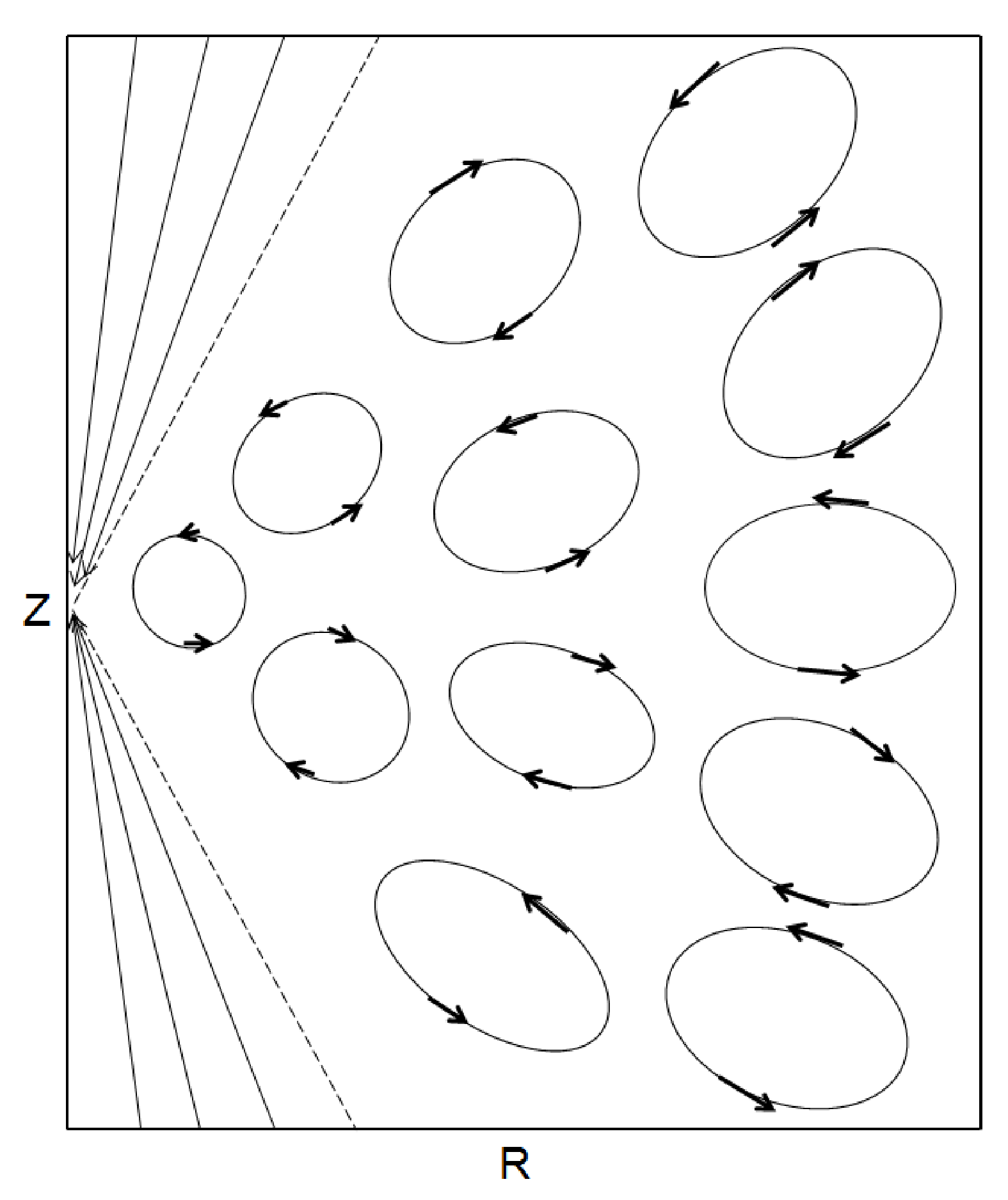}
\caption{Schematic figure of the flow motion in a CAAF. Close to the poles, there is
systematic inflow. Away from the poles, the flow is convection
dominated. The inward decrease of inflow rate is because the
convective mass flux decrease inward. \label{Fig:cartoon}}
\end{figure}

If the wind is absent, what is the reason for the inward decrease of inflow rate? Guided by our numerical simulation data, we draw a schematic figure of the motion of the flow in Figure {\ref{Fig:cartoon}}. We find that systematic real inflow occurs mainly close to the poles. Away from
the poles, the flow is full of convective eddies.  This reminds us the
scenario of convection-dominated accretion flow (CDAF) proposed by
Narayan et al. (2000) and Quataert \& Gruzinov (2000), although
their model was proposed to explain the dynamics of BHAFs rather
than CAAFs\footnote{YBW12 and Narayan et al. (2012) have shown that
the CDAF scenario may be not applicable to BHAFs.}.

Following Narayan et al. (2000) we now try to understand
quantitatively why the inflow rate decreases inward. The mass inflow
rate can be estimated by adding up all the inflowing gas elements.
This would give $\dot {M}_{\rm in} \sim 4\pi \rho \omega RH/2$, with
$R$ being the radius in cylindrical coordinates, $H$ is the half
thickness of the flow, $\omega$ is the rms velocity of the turbulent
eddy. Narayan et al. (2000) shows that $\omega \propto v_k$, where
$v_k$ is the Keplerian velocity of the flow. In model A1, we found
that $\rho \propto r^{-0.5}$. For the specific gravitational
potential we adopt, the Keplerian velocity outside and inside of
$r=0.1$ follows $v_k \propto r^0$ and $v_k \propto r^{-0.5}$,
respectively. Therefore, $\dot {M}_{\rm in} \propto \rho r^2 \omega
\propto r$ or $r^{\frac{3}{2}}$, i.e., the mass inflow rate
increases outward. Of course, such a rough estimation can't explain
quantitatively the simulation result of eq. 13.


\subsection{Dependence on model parameters}
\subsubsection{The form of shear stress}

\begin{figure}
\includegraphics[width=8cm]{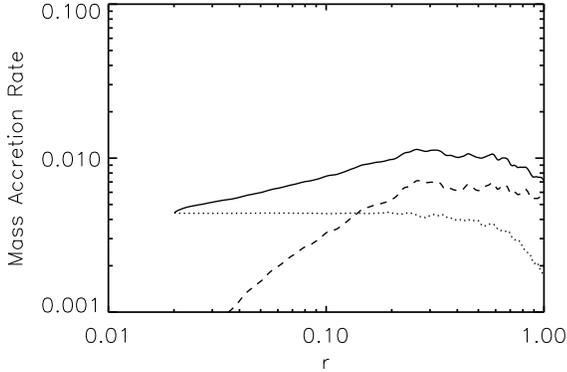}
\caption{The radial profile of the time-averaged (from $t=21$ to 26
orbits) and angle integrated mass inflow rate $\dot{M}_{\rm in}$
(solid line), outflow rate $\dot{M}_{\rm out}$ (dashed line), and
the net rate $\dot{M}_{\rm acc}$ (dotted line) in model A3.  \label{Fig:accretionA3}}
\end{figure}

In order to study whether the results depend on the form of shear
stress, we have carried out model A3. The only difference between
models A1 and A3 is the form of shear stress. In model A3, we assume
the shear stress $\nu \propto r$.
For model A3, we also find that the flow is convectively unstable.
The snapshot of density and entropy of model A3 is very similar to
that in model A1. Also, we find that the
mass inflow rate is a function of radius. Figure
\ref{Fig:accretionA3} shows the mass accretion rates for model A3.
The radial profile of the inflow rate from $r=0.02$ to $0.4$ can be
described by
\begin{equation}
\dot M_{\rm in} \propto r^{0.4} \end {equation}

From Figure \ref{Fig:accretionA3}, we can see  a significant mass
outflow rate. In order to study whether the outflow rate is
dominated by real outflow, we have also plotted the snapshot of
velocity vector for model A3. The result is very similar to that of
model A1, i.e., full of convective eddies. Trajectory analysis have
also been performed to study whether there are strong winds. We find
at $r=0.2, 0.4$ , the ratio of mass flux of winds to the total
outflow rate calculated by equation (\ref{outflowrate}) is $0.1\%,
0.2\%$, respectively. Again, there is almost no real outflows.


\begin{figure}
\includegraphics[width=8cm]{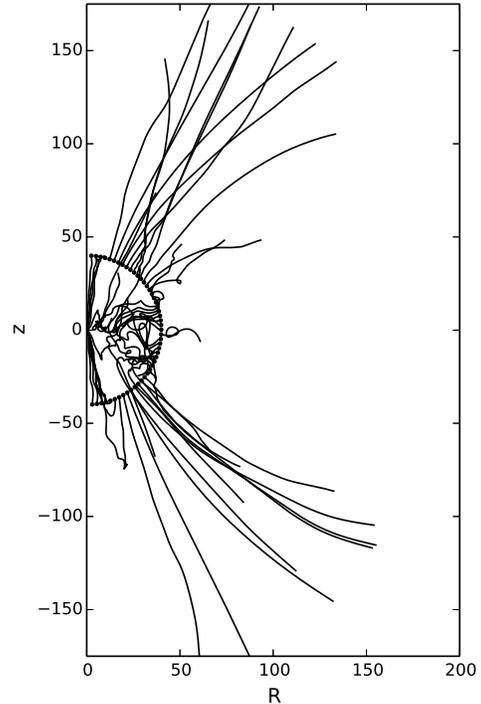}
\caption{Trajectory of gas for model B. The black dots located at
$r=40r_s$ are the starting points of the test particles. Significant
winds are clearly present. The horizontal and vertical axis are in
unit of $r_s$. \label{Fig:trajectoryB}}
\end{figure}

\begin{figure}
\includegraphics[width=8cm]{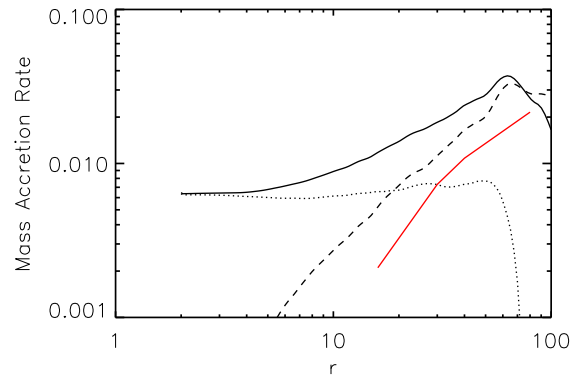}
\caption{The radial profile of the time-averaged (from $t=1.5$ to 2 orbits) and angle integrated mass inflow rate $\dot{M}_{\rm in}$
(solid line), outflow rate $\dot{M}_{\rm out}$ (dashed line), and
the net rate $\dot{M}_{\rm acc}$ (dotted line) in model B. The red line corresponds to the mass
flux of wind. The horizontal axis is in unit of $r_s$.
 \label{Fig:accretionB}}
\end{figure}

\subsubsection {Resolution}

In order to test whether the results obtained in this paper depend
on resolution, we have carried out model A2. In model A2, the number
of grids is $668 \times 320$, the resolution is two times higher
than that of model A1.

In model A2, the flow is still convectively unstable. Again, we find
that the mass outflow rate is dominated by convective outflows.
Using the trajectory method, we find that the ratio between the real
outflows to the total outflows is $\sim 0.1\%$.

\subsubsection {The gravitational potential}
Previous works have shown that strong winds are present in BHAFs
(e.g. YBW12; Li et al. 2013)\footnote{Narayan et al. (2012) find
much weaker wind. The discrepancy between YBW12 and Narayan et al.
(2012) was analyzed in Yuan et al. (2015).}. In this paper, we have
also performed simulations of BHAF. In model B, we simulate the BHAF
very close to the black hole, from $2$ to $400r_s$. For such a
region so close to the black hole, the gravity of the nuclear star
cluster can be neglected.

We have performed the trajectory analysis for model B. The results
are shown in Figure \ref{Fig:trajectoryB}. Clearly, significant
winds are present. Figure \ref{Fig:accretionB} shows the profile of
mass accretion rate. The red line in the figure corresponds to the
mass flux of winds. We find that this flux is approximately $70\%$
of the total mass outflow rate calculated from eq. (11). In Yuan et
al. (2015), we have studied the MHD BHAFs around a Schwarzschild
black hole and found that the ratio between the wind flux to the
total outflow rate is $\sim 60\%$. So the results of HD and MHD
simulations are similar. From $40-80r_s$, the radial profile of the
real outflow rate (wind mass flux) can be described as \begin {equation} \dot M_{\rm
real} \propto r^{0.95} \end {equation} In Yuan et al. (2015), the
radial profile of the real outflow mass flux is $\dot M_{\rm real}
\propto r$. The consistency between the present HD simulation and
the MHD simulation of Yuan et al. (2015) supports our HD simulation
study of wind as a good approximation of the more realistic MHD
study. The underlying physical reason of the consistency between
them is discussed in Begelman (2012).


\section{CONCLUSION AND DISCUSSION}\label{sec:summary}
In previous works, it has been shown that strong winds exist in
black hole hot accretion flow (e.g., YBW12). The mass flux of wind follows
$\dot{M}_{\rm wind}=\dot{M}_{\rm BH}(r/20r_s)$ (Yuan et al. 2015). In this paper, we have investigated
what the value of $r$ can be, i.e., whether or where the wind production will
stop., by performing hydrodynamical simulations. The key
difference between the present work and the previous ones is that we
must include the gravitational potential of the nuclear star cluster
in addition to the black hole potential at such large radii.
We call the accretion flow in the two regions where the
potential is dominated by the nuclear star cluster or by the black hole potential
``circum-AGN accretion flow'' (CAAF) and ``black hole accretion
flow'' (BHAF), respectively.  We find that same as BHAFs, CAAFs
are also convectively unstable and the inflow rate is also a power
law function of radius. However, different from a BHAF, there is
almost no wind in a CAAF.  For CAAFs, the
inward decrease of inflow rate is not because of strong wind, which is the case of a BHAF, but
because the motion is dominated by convection. This scenario is similar to
the convection-dominated accretion flow (CDAF) proposed by Narayan
et al. (2000) and Quataert \& Gruzinov (2000), although the CDAF model
is proposed to describe BHAFs rather than CAAFs.

This result indicates that the wind production stops at the radius
$R_A$ (refer to eq. 5), where the gravity of the black hole and that
of the nuclear star cluster equal to each other. Combining it with
the result of Yuan et al. (2015), the mass flux of wind is described
by:
\begin{equation}\dot{M}_{\rm wind}=\dot{M}_{\rm BH}(r/20r_s), ~~r\la R_A.\end{equation} $\dot{M}_{\rm wind}=\dot{M}_{\rm BH}(R_A/20r_s)$ when $r\ga R_A$. It is interesting to compare $R_A$ with the Bondi radius, $R_B(\equiv GM_{\rm BH}/c_s^2$), which is generally regarded as the outer boundary of the accretion flow.  Here $c_s$ is the sound speed of the accreting gas at the Bondi radius. Its typical value is $c_s\equiv \sqrt{\gamma p/\rho}\sim  300 (T_{\rm ISM}/10^7K)^{1/2} {\rm km~s^{-1}}$, with $T_{\rm ISM}$ being the temperature of the interstellar medium around the Bondi radius. This is very close to the typical value of $\sigma$, which is $\sim (100-400) {\rm km~s^{-1}}$ (e.g., Kormendy \& Ho 2013). So in practice, we can assume that wind stops to be produced outside of the outer boundary of the accretion flow.

The disappearance of winds in CAAFs must be due to the change of the gravitational potential. Specifically, we speculate that it is very likely because of the
change of the slope rather than the absolute value of the gravitational potential. In this context we note that the change of differential rotation of the accretion flow, which could be due to the change of the slope of the gravitational potential, causes the change of magnetorotational instability (e.g., Pessah et al. 2008; Penna et al. 2013).  But the exact reason still remains to be probed.

\section*{Acknowledgements}

We thank Ramesh Narayan for helpful discussions. This work was
supported in part by  the National Basic Research
Program of China (973 Program, grant 2014CB845800), the
Strategic Priority Research Program ¡°The Emergence of Cosmological
Structures¡± of the Chinese Academy of Sciences (grant XDB09000000), and the Natural Science Foundation of China (grants 11103061, 11133005, 11121062, and 11573051).
This work has made use of the High Performance Computing Resource in the
Core Facility for Advanced Research Computing at Shanghai
Astronomical Observatory.

\label{lastpage}

\end{document}